\documentstyle[12pt]{article}
\newcommand{\beq}{\begin{equation}}
\newcommand{\eeq}{\end{equation}}
\newcommand{\bea}{\begin{eqnarray}}
\newcommand{\eea}{\end{eqnarray}}

\newcommand{\rfn}[1]{(\ref{#1})}
\newcommand{\Eq}[1]{Eq.~(\ref{#1})}

\renewcommand{\titlepage}{\clearpage%
\setcounter{footnote}{0}%
\thispagestyle{empty}\pagestyle{plain}\pagenumbering{arabic}%
\kern1mm
\vskip15mm\normalsize}
\newcommand{\docnum}[1]{\hbox to \hsize{\hskip123mm\hbox{#1}\hss}}
\renewcommand{\date}[1]{\hbox to \hsize{\hskip123mm\hbox{#1}\hss}}

\renewcommand{\title}[1]{\vskip1em\begin{center}\Large\bf#1\end{center}\vskip2.5em}
\renewcommand{\author}[1]{\vskip0.5em{\bf #1}\vskip0.5em}
\renewcommand{\abstract}{\begin{center}{\Large \bf
Abstract}\end{center}\quotation}

\begin{document}
\def\thefootnote{\fnsymbol{footnote}}
\begin{titlepage}
\title
{On the Interpretation of the Electroweak Precision Data
\footnote[2]{\normalsize
Partially supported by Deutsche Forschungsgemeinschaft
\hfil\break
\footnotesize
A preliminary version of this paper was presented by D. Schildknecht
at the Marseille International Conference, July 1993}}
\begin{center}
\author{M.~Bilenky\footnote[1]
{\normalsize  Alexander von Humboldt Fellow, on leave from
JINR, Dubna, Russia.},
{}~K. Kolodziej\footnote[3]
{\normalsize  Permanent address: Dept. of Physics, University of Silesia,
Katowice, Poland},
{}~M. Kuroda\footnote[4]
{\normalsize  Permanent address: Dept. of Physics, Meiji-Gakuin University,
Yokohama, Japan}
{}~~and~~
D. Schildknecht}
\end{center}

\vskip0.3em
{\it Department of Theoretical Physics, University of Bielefeld,
33501 Bielefeld, Germany}
\vskip0.5em

\vspace{0.5cm}
\begin{abstract}
The recent precision electroweak data on $\Gamma^l, \bar s^2_W$ and
$M_W/M_Z$ are compared with the tree-level and the dominant-fermion-loop
as well as the full one-loop standard-model predictions. While
the tree-level predictions are ruled out, the dominant-fermion-loop
predictions, defined by using $\alpha (M^2_Z)\cong 1/128.9$ in the
tree-level formulae, as well as the full one-loop predictions are
consistent with the experimental data. Deviations from the
dominant-fermion-loop predictions are quantified in terms of an effective
Lagrangian containing three additional parameters which have a
simple meaning in terms of $SU(2)$ symmetry violation.
The effective Lagrangian yields the standard  one-loop predictions
for specific values of these parameters, which are determined by $m_t$
and $m_H$.
\end{abstract}
\end{titlepage}
\setcounter{page}{1}
\def\thefootnote{\arabic{footnote}}
\setcounter{footnote}{0}
\setcounter{section}{1}
To start with, in figs. 1a, b, c, we show the three projections of
the three-dimensional, 68 \% C.L. volume defined by the data in
$(M_W/M_Z, \bar s^2_W, \Gamma_l)$-space in comparison with the simple-minded
standard-model \cite{GWS} tree-level prediction obtained from \footnote{We use
the standard notation, $M_W$ and $M_Z$ for the masses of the charged
and neutral weak boson, respectively, $\bar s^2_W$ for the square
of the weak leptonic mixing angle measured at $\sqrt s = M_Z$ and
$\Gamma_l$ for the leptonic width of the $Z_0$, etc.}
\footnote{We confine our analysis to the mentioned observables,
as these are particularly simple ones which do not involve important
hadronic (gluonic) effects.}
\bea
\bar s^2_W (1-\bar s^2_W) & = & {{\pi \alpha (0)}\over {\sqrt 2 G_\mu
M^2_Z}},\nonumber\\
{{M^2_{W^{\pm}}}\over{M^2_Z}} & = & 1 - \bar s^2_W, \label{1}\\
\Gamma_l & = & {{G_\mu M^3_Z}\over {24 \pi \sqrt 2}} \bigl( 1 +(1-4 \bar s^2_W
)^2\bigr)\nonumber
\eea
by using
\bea
\alpha(0) & = & 1/137.0359895(61)\nonumber\\
G_\mu & = & 1.16639(2) 10^{-5} GeV^{-2}\label{2}\\
M_Z & = & 91.187 \pm 0.007 GeV\nonumber
\eea
as input parameters. The data on $\bar s^2_W$ and $\Gamma_l$ are the
average values of the results of the four LEP experiments
presented at the Marseille conference \cite{LEF,ALT}.
The value of $\bar s^2_W$ is obtained from the ratio of the vector and
axial-vector couplings of the $Z_0$ to leptons,
$g_V/g_A$, via
\beq
\bar s^2_W = {1\over 4} \bigl( 1 - {{g_V}\over {g_A}}\bigr), \label{3}
\eeq
where $g_V/g_A$ is deduced from all asymmetries at $\sqrt s = M_Z$,
\bea
g_V/g_A{\rm(all~asymmetries)} & = & 0.0712 \pm 0.0028,\nonumber\\
\bar s^2_W & = & 0.2322 \pm 0.0007.\label{4}
\eea
The leptonic width measured at LEP is given by
\beq
\Gamma_l = 83.79 \pm 0.28 MeV,\label{5}
\eeq
where lepton universality is assumed, and the ratio
\beq
{{M_W}\over{M_Z}} = 0.8798 \pm 0.0028 \label{6}
\eeq
is given by CDF and UA2 data \cite{CDF}
\par
As seen from figs. 1a, b, c the tree-level prediction \rfn{1} is clearly ruled
out by the data.
\par
\medskip
In 1988, as a strategy for the analysis of $Z_0$ precision data, it was
suggested \cite{GS} "to isolate and to test directly the "new physics" of
boson loops and other new phenomena by
comparing with and looking for deviations from the predictions
of the dominant-fermion-loop calculations". The dominant-fermion-loop
predictions are simply obtained by the replacement\footnote{
In ref. \cite{GS}, the term "dominant-fermion-loop" includes the
loop contribution of a top quark of mass $30 GeV \leq m_t \leq
200 GeV$. Compare note added in proof for a discussion of
the top-quark effect.}
\beq
\alpha(0) \to \alpha(M^2_Z) = 1 / 128.87 \pm 0.12\label{7}
\eeq
in \rfn{1}. The replacement \rfn{7} takes care of the dominant radiative
corrections due to the logarithms of widely different scales
(masses of the light fermions and the $Z^\circ$ mass). The correction
due to the replacement \rfn{7} is precisely calculable as a pure QED
\footnote{Strictly speaking, this holds for lepton loops. The
vacuum polarization due to quarks requires calculations based on the
data on $e^+e^-\to$ hadrons via dispersion relations in order to
incorporate hadronic (QCD) corrections. The uncertainties in the
input data lead to the error in \rfn{7}. Compare ref. \cite{BJPZ}.}
effect, and, consequently, it is independent of any hypothesis on the
"new physics" (i.e., the empirically untested physics)
of the interactions of the vector bosons with each other
entering bosonic loop corrections and
the properties of the Higgs particle or, from the point of view
of local gauge symmetry,
the nature of the electroweak symmetry breaking.
\par
Accordingly, in our second step, in figs. 2a,b,c, we take a
closer look at the data contours and compare them with predictions of
the dominant-fermion-loop calculations (indicated by the symbol "star"
in figs. 2a,b,c) and the full one-loop predictions thus
refining and extending a previous analysis \cite{KKS} based on the $^\prime$89
LEP data. The lines in figures 2a,b,c, give the predictions of the
one-loop results for fixed values of $m_H = 100, 300$ and
$1000 GeV$, treating $m_{top}$ as a parameter as indicated. These
predictions were calculated by using the ZFITTER program \cite{BAR}. They are
in agreement with calculations based on refs. \cite{KKS,KMS}.
{}From figs. 2a,b,c, we conclude:
\begin{enumerate}
\item
The data at 68 \% C.L. are consistent with the
dominant-fermion-loop results.\footnote{This was also noted and
strongly emphasized by Novikov, Okun and Vysotsky \cite{NOV}, who showed
in addition \cite{NOV2} that the experimental results for the LEP observables
involving hadronic $Z^0$ decays are consistent with the $\alpha (M^2_Z)$
tree-level formula. Compare also ref. \cite{ABC}.
Consistency of the data with both, the
dominant-fermion-loop as well as the full one-loop theoretical
predictions is not entirely unexpected, as it was repeatedly
stressed \cite{GS,KKS} that the experimental errors to be expected in the
high-precision LEP measurements will at most marginally
allow to discriminate
between dominant-fermion-loop predictions and full one-loop results.}
\item
The data put a strong bound on deviations from the
dominant-fermion-loop prediction, independently of the origin
(standard or non-standard) of such additional contributions.
\item
The data are in agreement with the full one-loop standard-model
predictions
which extend our present
empirical knowledge by introducing non-Abelian bosonic self-interactions,
at the same time assuring renormalizability of the theory at
the expense of postulating the existence of the Higgs scalar particle
(apart from introducing the top quark).
\end{enumerate}
\par
Evidently, apart from discovering the top quark and determining
its mass, a direct measurement of the (trilinear and even
quadrilinear) couplings of the vector bosons to each other \cite{BKRS} and a
discovery of the Higgs particle seems indispensable for a full
verification of the present electroweak theory.
\par
The agreement between theory and experiment in figs. 2a,b,c is
(obviously) far from trivial. This is best appreciated by reminding
oneself of the pre-UA1,-UA2 and pre-LEP era. In 1978, it was
pointed out by Bjorken and by Hung and Sakurai \cite{HS} that the
experimental data on neutral-current interactions then available
did not necessarily require the validity of the $SU(2)_L \times U
(1)_Y$ standard theory. It was shown that a Lagrangian
based on global $SU(2)$ weak-isospin
symmetry broken by $\gamma W^3$ mixing could also explain
the NC data, and it was stressed that high-energy (LEP, SLC)
data were necessary to rule out this four-free-parameter
effective-Lagrangian alternative.
The high-precision data available now allow one
to precisely investigate to what degree of accuracy any
$SU(2)_L\times U(1)_Y$ violation of the Hung-Sakurai type is
actually excluded.
\par
Somewhat more generally, we propose to analyse the body of experimental
data on $M_W/M_Z, \bar s^2_W$ and $\Gamma_l$ on
the basis of the effective Lagrangian (written in the
physical base)
\bea
L_C &=&
-{1\over 2} W^{\mu \nu}_+ W^-_{\mu \nu} +
{{g_{W^\pm}}\over {\sqrt 2}}
(j_\mu^+ W_+^\mu + h. c.) + M^2_{W^\pm} W_\mu^+ W_-^\mu, \nonumber\\
L_N &=&
-{1\over 4} A^{\mu \nu} A_{\mu \nu} -
{1\over 4} Z^{\mu \nu}_0 Z^0_{\mu \nu} +
{1\over 2}{{M^2_{W^0}}\over{1-\bar s^2_W(1-\epsilon)}}
Z^{\mu}_0 Z^0_{\mu}\label{8}\\
&~&+e J^\mu_{em} A_\mu + {{g_{W^0}}\over
{\sqrt{1-\bar s^2_W(1-\epsilon)}}}
\bigl( J^\mu_3 - \bar s^2_W J^\mu_{em}\bigr) Z^0_\mu,\nonumber
\eea
with the constraint
\beq
g^2_{W^0} = {{e^2}\over{\bar s^2_W}}(1-\epsilon),
\label{9}
\eeq
by which the number of parameters is reduced to {\it six} independent
ones. The parameter $e$ in \rfn{8} refers to the electric charge measured
at the scale $M_Z, e=e(M^2_Z)$. The Lagrangian \rfn{8} contains $SU(2)_L$
violation via mixing \`a la Hung-Sakurai \footnote{This is explicitly
seen by performing appropriate transformations to the $BW^3$ or to the
$\gamma W^3$ base. Compare refs. \cite{KKS,BKRS2,GKS}.}
\beq
\epsilon W^{3\mu\nu} B_{\mu\nu},
\label{10}
\eeq
quantified by the parameter $\epsilon$ and $SU(2)$ violation
via discriminating
$g_{W^\pm}$ from $g_{W^0}$, quantified by the parameter $y$, in
\bea
g^2_{W^\pm} &=&y g^2_{W^0},\nonumber\\
&=&(1+\Delta y) g^2_{W^0}.
\label{11}
\eea
A violation of global $SU(2)$ symmetry in the mass term is introduced
and quantified by discriminating $M_{W^\pm}$ from $M_{W^0}$ in
\bea
M^2_{W^\pm} &=&x M^2_{W^0},\nonumber\\
&=&(1+\Delta x)M^2_{W^0}.
\label{12}
\eea
Altogether, the standard number of {\it three} input parameters
is thus supplemented by {\it three} parameters quantifying {\it all}
potential sources of $SU(2)$ violation. We note that any additional
violation of $SU(2)$ symmetry can be removed
if a redefinition of the fields is accompanied by a readjustment
of the $SU(2)$-violating parameters introduced in
\rfn{10},\rfn{11},\rfn{12}.
\par
A further remark on the significance of Lagrangian \rfn{8} may be
appropriate. It contains a twofold physical interpretation:
\begin{enumerate}
\item
The Lagrangian \rfn{8} effectively describes the one-loop standard-model
electroweak interactions of the leptons \footnote{For quarks, \rfn{8}
has to be generalized in a manner such that the large
non-universal corrections due
to $Z\to b\bar b$ decay \cite{ABR} can be accomodated.}
with the vector bosons at
the $Z^0(W^\pm)$ mass scale, provided $x, y$ and $\epsilon$ are
appropriately specified in terms of $m_t$ and $m_H$ (see below).
\item
Without employing the concept of spontaneous symmetry breaking,
the effective Lagrangian \rfn{8} for $x=y=1$ but $\epsilon \not= 0$ is
obtained by assuming breaking of global $SU(2)$ symmetry via current
mixing \cite{HS}. A priori, from this point of view, $\epsilon$ may differ
appreciably from $\epsilon=0$. The strong empirical constraints
on $\epsilon$ to be given below reduce $\epsilon$ to the order
of magnitude of standard radiative corrections and strongly restrict
appreciable non-standard effects, thus empirically verifying the
"unification condition" \cite{HS} $\epsilon \cong 0$ to a high level of
accuracy. The additional parameters $\Delta x \not= 0, \Delta y \not= 0$
take into account additional potential sources of $SU(2)$ violation.
\end{enumerate}
\medskip
On the basis of the effective Lagrangian \rfn{8}, the modified
relations \rfn{1} now read
\bea
\bar s^2_W (1-\bar s^2_W) & = & {{\pi \alpha (M_Z)}\over {\sqrt 2 G_\mu
M^2_Z}} {y \over x} (1-\epsilon)
{1 \over {(1+{{\bar s^2_W} \over {1-\bar s^2_W}} \epsilon)}},\nonumber\\
{{M^2_{W^{\pm}}}\over{M^2_Z}} & = & (1 - \bar s^2_W) x
(1+{{\bar s^2_W} \over {1-\bar s^2_W}} \epsilon),
 \label{13}\\
\Gamma_l & = & {{G_\mu M^3_Z}\over {24 \pi \sqrt 2}}
\bigl( 1 +(1-4 \bar s^2_W
)^2\bigr) {x \over y} (1+ {{3\alpha} \over {4 \pi}}),\nonumber
\eea
where a QED correction factor is included in the expression
for $\Gamma_l$ in agreement with the definition of $\Gamma_l$ used
in the analysis of the data \cite{LEF,ALT}.
Keeping only terms linear in $\epsilon, \Delta x, \Delta y$, one
obtains from \rfn{13}\footnote{It was checked that the {\it relative}
errors
in $\bar s^2_W, M_W/M_Z$ and $\Gamma_l$ introduced by the linear
approximation are of the order of $10^{-4}$ for the relevant range
of values of $\epsilon, \Delta x$, and $\Delta y$.}
\bea
\bar s^2_W &=&s^2_0\left [ 1 - {1\over{c^2_0 - s^2_0}} \epsilon
- {{c^2_0}\over{c^2_0 - s^2_0}} (\Delta x - \Delta y)\right ],\nonumber\\
{{M_W}\over {M_Z}}
&=&c_0\left [ 1 + {{s^2_0}\over{c^2_0 - s^2_0}} \epsilon
+ {{c^2_0}\over{2(c^2_0 - s^2_0)}} (\Delta x - \Delta y) +
{1\over 2} \Delta y \right ],\label{14}\\
\Gamma_l &=&\Gamma_l^{(0)}
\left[ 1\! +\!
{{8 s^2_0(1-4 s^2_o)}\over
{(c^2_0\! -\! s^2_0)(1\!+\!(1\!-\!4 s^2_0)^2)}}
\epsilon\! +\!
{{2(c^2_0 - s^2_0 - 4 s^4_0)}\over
{(c^2_0\! -\! s^2_0)(1\!+\!(1\!-\!4 s^2_0)^2)}}
(\Delta x\! -\! \Delta y)\right],
\nonumber
\eea
where
\bea
& s^2_0 (1-s^2_0)\equiv c^2_0 s^2_0 =
{{\pi \alpha (M^2_Z)}\over
{\sqrt 2 G_\mu M^2_Z}},
\label{15}\\
\Gamma^{(0)}_l = &
{{\alpha (M^2_Z)M_Z}\over
{48 s^2_0 c^2_0}} \left[ 1 + (1-4 s^2_0)^2\right]
\left( 1+{{3\alpha}\over
{4\pi}}\right).\nonumber
\eea
In view of future analysis of empirical data, it will be useful
to explicitly give the inversion of \rfn{14},
\newpage
\bea
\epsilon & = &
{1\over{(c^2_0 - s^2_0)^2 + 4 s^4_0}}\nonumber\\
&&\left[ -
{{c^2_0 - s^2_0 - 4 s^4_0}\over {s^2_0}} \bar s^2_W -
{{c^2_0 (1 +(1-4 s^2_0)^2)}\over{2\Gamma_l^{(0)}}}
\Gamma_l + 2 c^2_0 - 5 s^2_0 + 8 c^2_0 s^4_0 \right],\nonumber\\
\Delta x - \Delta y & = &
{1\over {(c^2_0 - s^2_0)^2 + 4 s^4_0}}
\left[ 4 (1-4 s^2_0) \bar s^2_W +
{{1+(1-4 s^2_0)^2}\over {2 \Gamma _l^{(0)}}}
\Gamma_l - 1 + 8 s^4_0\right],\label{16}\\
\Delta y & = &
{{-2}\over {(c^2_0 - s^2_0)^2 + 4 s^4_0}}\nonumber\\
&&\left[(c^2_0\! -\!5 s^2_0) \bar s^2_W\! +\!
{{c^2_0(1\!+\!(1\!-\!4 s^2_0)^2)}\over {4 \Gamma _l^{(0)}}}
\Gamma_l\! -\! {1\over 2}\! +\! {3\over 2} s^2_0\! +\! 4 s^6_0\right]
\!+\! 2 \left( {{M_W}\over{M_Z c_0}}\! -\! 1\right).\nonumber
\eea
The dominant-fermion-loop prediction \rfn{7} in the Lagrangian \rfn{8}
(with \rfn{9}, \rfn{11}, \rfn{12})
simply corresponds
to $\epsilon = \Delta x = \Delta y = 0$,
while the one-loop induced standard-model corrections are described
by Lagrangian \rfn{8} for special values of $\epsilon, x, y$, which
depend on the top-mass, $m_t$, and the Higgs mass, $m_H$. Using the
results of, e.g., \cite{GS}, one finds for the dominant contributions
to $\Delta x - \Delta y, \Delta y$ and $\epsilon$
\bea
\Delta x - \Delta y
&=&{{3 G_\mu m^2_t}\over {8\sqrt 2 \pi^2}} -
{{G_\mu M^2_W}\over {2\pi^2\sqrt2}} {3\over 2} {{s^2_0}\over
{c^2_0}} \ln {{m_H}\over {M_Z}} + \cdots ,\nonumber\\
\Delta y &=&
{{G_\mu M^2_W}\over {2\sqrt2\pi^2}}
\ln {{m_t}\over {M_Z}} + \cdots ,\label{17}\\
\epsilon &=&
{{G_\mu M^2_W}\over {6\pi^2\sqrt 2}}\ln {{m_t}\over{M_Z}} -
{{G_\mu M^2_Zc^2_0}\over {12\pi^2\sqrt2}}
\ln {{m_H}\over {M_Z}} + \cdots .\nonumber
\eea
Comparison with ref. \cite{ABC} reveals that our parameters defined in
terms of $SU(2)$-symmetry properties of electroweak interactions
are simple
linear combinations of the parameters $\epsilon_{N1},
\epsilon_{N2}, \epsilon_
{N3}$ introduced in \cite{ABC}\footnote{For related
work based on parameters somewhat different from the ones in \rfn{18},
we refer to the list of references in \cite{ABC}.}
 by the requirement of isolating the
quadratic $m_t$ dependence, i.e.,
\bea
\Delta x - \Delta y & = & \epsilon_{N1},\nonumber\\
\Delta y & = & - \epsilon_{N2},\label{18}\\
\epsilon & = & - \epsilon_{N3}.\nonumber
\eea
Relations \rfn{18} establish the meaning of the parameters $\epsilon_{N1},
\epsilon_{N2}, \epsilon_{N3}$ with respect to $SU(2)$-symmetry
properties of the electroweak interactions.
\par
By solving \rfn{13} for $\epsilon, \Delta x - \Delta y, \Delta y$
and expanding around the experimental values for
$M_W/M_Z, \bar s^2_W, \Gamma_l$ or, alternatively, by using \rfn{16},
one obtains linear relations for $\epsilon, \Delta x - \Delta y,
\Delta y$ in terms of
$M_W/M_Z, \bar s^2_W, \Gamma_l$.
By inserting the experimental data, one can deduce the experimental
bounds on
$\epsilon, \Delta x - \Delta y,$ and $\Delta y$.
\par
In figs. 3a, b, c, we show the resulting projections of the 68 \% C.L.
volume in
($\epsilon, \Delta y, \Delta x-\Delta y$)-space on the planes
$\Delta y = const, \epsilon = const$ and $\Delta x - \Delta y = const$,
respectively, in comparison with the $\alpha (M^2_Z)$-tree-level point
$(\epsilon = \Delta x - \Delta y = \Delta y = 0)$ and the
one-loop-standard-model predictions. As expected from figs. 2a,b,c,
the $SU(2)$-symmetry point\footnote{The fact that the point
$\epsilon = \Delta y = 0$ in fig. 3c lies at the outside edge of
the ellipse (in distinction from the corresponding points in figs. 2 a,b,c)
is related to the linearization procedure applied to \rfn{12} (and also used
in \rfn{13}) when deriving the covariance matrix from the experimental data.},
$\epsilon = \Delta x - \Delta y = \Delta y = 0$ as well as the
one-loop predictions lie within the ellipsoid. As a new result from
figs. 3a, b, c, in comparison with figs. 2a, b, c, one can read off
the restrictions on the magnitude of the departure from $SU(2)$
symmetry allowed by the experimental data. The absolute values of
$\epsilon, \Delta x - \Delta y,$ and $\Delta y$
at 68 \% C.L. are restricted to the level of (10 to 20) $\cdot$
10$^{-3}$ or 1 to 2 per cent.
\par
As seen in Figs. 3a,b,c, the one-loop standard model for
$m_t \buildrel >\over\sim 110 GeV$ yields approximately constant
values of $\epsilon \cong - 6.5 \cdot 10^{-3}$ and
$\Delta y \cong + 6.5 \cdot 10^{-3}$, while, as a consequence
of the quadratic $m_t$ dependence, the quantity $\Delta x - \Delta y$
varies significantly, when $m_t$ is varied between $100$ and $200 GeV$.
\par
In conclusion, the high precision of the present data is apparent
from the fact that the tree-level prediction based on the "wrong" value
of $\alpha \equiv \alpha(0) \equiv 1/137$ is clearly ruled out,
while the tree-level prediction obtained by inserting the "correct"
value of $\alpha \equiv \alpha(M^2_Z) \cong 1/129$ is consistent
with the experimental data. Deviations from the prediction based on
$\alpha (M^2_Z)$ are indeed strongly constrained by the data.
The parameters $\epsilon, \Delta x, \Delta y$ quantifying $SU(2)$-symmetry
violations are constrained to the order of $0.01$ to $0.02$ at 68 \% C.L.
This is the order of magnitude of the standard one-loop weak corrections
which depend on the magnitude of the mass of the top quark, the empirically
unknown interactions of the vector bosons among each other and the
existence of the Higgs scalar. The consistency between the data
and the standard one-loop predictions strongly supports the validity
of the standard theory, even though, in principle, any theory which
supplies sufficiently small bosonic loops (assuming the existence
of the top quark with a reasonable value of its mass) will be consistent
with the available experimental information. By the end of the running
of LEP at the $Z^0$ energy, the precision of the data can be envisaged
to become even better than the distance between the
dominant-fermion-loop point and the standard-model lines in figs. 2 and 3.
Even though constraints on non-standard effects will become even
stronger than at present, a
direct investigation of the
trilinear (and quadrilinear) couplings among
the vector bosons and direct
experimental evidence for the Higgs scalar will be indispensable for a
full empirical verification of our present theory of the electroweak
interactions.
\vskip 1 truecm
\leftline{\underbar{Note added in proof.}}
\medskip
The dominant-fermion-loop analysis of the present work did not
take into account the direct experimental information on the
properties of a (hypothetical) top quark. In particular, the
lower bound on the mass, $m_t \geq 112 GeV$
\cite{BG} of a standard top quark was not taken into account.
Even though the top quark has not been identified experimentally
so far, strong arguments are available for its existence.
Let us thus assume that a top quark of mass $m_t \geq
112 GeV$ exists and moreover that
it has standard interactions with the
electroweak vector bosons. As a consequence, the top-quark loop
yields a well-known contribution to the electroweak
observables which has to be taken into account
in addition \cite{GS,KKS}
to the running of the electromagnetic coupling
caused by the light leptons and quarks.
\par
Taking into account the standard one-loop top contribution
in addition to the contribution of the light leptons and
quarks yields the short-dashed and long-dashed curves in
figs. 2a,b,c and 3a,b,c. The short-dashed curve is calculated
by using the asymptotically leading (quadratic and logarithmic)
$m_t$ dependence given in \Eq{17}. The long-dashed curve takes
into account the one-loop top-quark contribution exactly.
\medskip
{}From figs. 2a,b,c and 3a,b,c, we conclude:
\smallskip\noindent
\begin{enumerate}
\item
The dominant-fermion-loop approximation, taking into account
a standard "light" top quark of mass $m_t\cong 60 GeV$ to
$m_t\cong 80 GeV$ in addition to the light fermions, is
marginally consistent with the LEP precision data (disregarding
the fact that the negative results of the direct searches
provide a lower bound of $m_t \geq 112 GeV$).
\item
The assumption of a standard top-loop contribution to the LEP
observables combined with the lower bound of $m_t \geq
112 GeV$ from the direct top search leads to a
(mild) discrepany between the dominant-fermion-loop predictions
and the LEP precision data at 68 \% C.L. (compare figs. 2a,b and,
in particular, fig. 3a). Consistency of the existence
of a very massive standard
top quark with the LEP data thus requires (standard or non-standard)
effects beyond the dominant fermion loops to contribute to the
LEP observables.
\end{enumerate}
\medskip
In summary, exploiting the lower bound of $m_t \geq
112 GeV$ for a standard top-quark obtained in
the direct top-quark search
allows us to refine the conclusion of the present paper:
consistency with the LEP precision data of the hypothesis of the
existence of a heavy top quark with mass $m_t \geq 112 GeV$ and
standard electroweak interactions, (marginally)
requires the presence of (standard or non-standard) contributions
to the LEP observables beyond the dominant fermion loops. In other
words, under the mentioned hypotheses on the top quark, LEP is
starting to "see" effects beyond the standard
dominant fermion loops.

\section*{Figure captions}

\begin{list}
{\bf Fig. 1:}
{\labelwidth1.5cm \leftmargin3cm \labelsep0.4cm}
\item
The experimental data for\hfil\break
a)
the boson-mass ratio, $M_M/M_Z$, and the electroweak mixing
angle,
$\bar s^2_W$, deduced from all the asymmetries measured at
$\sqrt s= M_Z$,\hfil\break
b)
the leptonic width, $\Gamma_l$, of the $Z^\circ$ and $\bar s^2_W$
\hfil\break
c)
$\Gamma_l$ and $M_W/M_Z$
\par
are compared with the standard tree-level predictions based
on $\alpha^{-1} \equiv \alpha (0)^{-1}\cong 137.036$. The
ellipses are the projections
of the 68 \% C.L.
ellipsoid
defined by the data in
$(M_W/M_Z, \bar s^2_W, \Gamma_l)$ space.
\end{list}
\par\noindent
\begin{list}
{\bf Fig. 2a,b,c:}
{\labelwidth2.5cm \leftmargin3cm \labelsep0.4cm}
\item

The same experimental data as in figs. 1a, b, c. The point denoted by
the symbol
"star" gives the standard tree-level predictions based
on $\alpha^{-1} \equiv \alpha (M^2_Z)^{-1} =
1/128.87\pm 0.12$.
The long-dashed curve is obtained by taking into account
the top-loop contribution (in addition to the light fermions
yielding $\alpha (M^2_Z)$).
The corresponding
short-dashed curve is calculated
by approximating the top
contribution according to \Eq{17}.
The standard-model full one-loop results
are shown for
Higgs-boson masses of $m_H = 100, 300$ and $1000 GeV$, varying
the mass of the top quark in steps of $20 GeV$, as indicated.
Note that the
error bar shown at the $\alpha (M^2_Z)$ point
denoted by the symbol "star" must also
be applied to all other
theoretical results shown in the figure.
\end{list}
\par\noindent
\begin{list}
{\bf Fig. 3a,b,c:}
{\labelwidth2.5cm \leftmargin3cm \labelsep0.4cm}
\item
The experimental restrictions (68\% C.L.) on the paramters
$\epsilon, \Delta x - \Delta y$ and
$\Delta y$ obtained by solving
\rfn{13} or, alternatively, by using \rfn{16} and evaluating
the result by
inserting the experimental data
for $M_W/M_Z, \bar s^2_W$ and $\Gamma_l$ given in
\rfn{4}, \rfn{5},
\rfn{6} and
displayed in figs. 1 and 2. The theoretical curves correspond
to
the ones of figs. 2a,b,c.
\end{list}
\end{document}